\documentclass [cits] {PoS}

\usepackage{amsmath}

\title{Oscillatory terms in the domain wall transfer matrix}

\ShortTitle{Oscillatory terms in the domain wall transfer matrix}

\author{Sergey Syritsyn and \speaker{John W. Negele} \\
\hbox{Center for Theoretical Physics, Massachusetts Institute of Technology, Cambridge MA, 02139 USA} \\
E-mail: \email{syritsyn@mit.edu, negele@mit.edu}}

\abstract{We study the transfer matrix for domain wall fermions to understand the origin 
and significance of oscillatory contributions to hadron correlation functions that 
arise for M >1. For a free particle in one space, one time, and one flavor 
dimension, the eigenmodes of the one-body operator appearing in the transfer 
matrix are calculated, and the role of the negative eigenmodes arising when M 
> 1 is studied.  In the case of three space dimensions, oscillatory behavior for hadron correlation functions in QCD is 
shown to emerge for free fermions when M exceeds 1, and to increase with increasing M. Analogous behavior is observed for domain wall fermions on HYP smeared MILC lattices, and a procedure is demonstrated for subtracting oscillating terms from physical observables.   
\bigskip

MIT-CTP 3875
}

\FullConference{The XXV International Symposium on Lattice Field Theory\\
		 July 30-4 August 2007\\
		 Regensburg, Germany}

\begin{document}

\section{Introduction}

Oscillatory terms in the dependence of lattice correlation functions on Euclidean time $t$ arise for domain wall fermions with domain wall mass $M >1$, so the goal of this work is to explore their origin, whether they are benign or malevolent, and removal of their contributions to physical observables.  

For domain wall fermions\cite{Kaplan:1992bt,Shamir:1993zy}, the domain wall mass $M$ generates a bound state in the extra  ``flavor'' direction $s$.  For free fermions, the region $ 0 < M < 2 $ is free of doublers, and the choice of $ M < 1 $ produces exponential decay $e^{-\alpha s} $ in the flavor direction.   The choice  $ M > 1 $ produces singularities in $\alpha$ and oscillating behavior for the propagator in the flavor direction\cite{Shamir:1993zy}  $G \sim (1-M)^{-s-s'}$.  For   fermions in a gauge field, an additive mass shift modifies the values $ M $ = 0, 1, and 2 above, and in practice, one uses spectral flow to determine the region of $M $  with the lowest density of low eigenmodes of the Hermitian Dirac operator, $ \gamma_5 D_W $.  Thus, for the DBW2 action\cite{Aoki:2002vt}, the LHPC hybrid action using DW fermions on HYP smeared MILC configurations\cite{Hagler:2007xi}, and the 2+1 dimensional DW action \cite{Antonio:2007tr}, the domain wall masses were determined to be $M$ = 1.7, 1.7, and 1.8 respectively, and in the first two cases, oscillatory terms in the $t$ dependence of hadron  correlations functions have been observed.

To understand the origin of the oscillatory modes, we study the behavior of the transfer matrix in $t$.  If one analyzes the lattice theory in the physical dimensions of space and $t$ by integrating out all the flavor slices except the boundaries containing the chiral fermions, the theory is intrinsically nonlocal and we are unable to construct a one-step transfer matrix.  However, we may alternatively consider  correlation functions in the full dimensions of space, $s$ and $t$, and restrict our attention to operators that are located on the flavor boundaries containing chiral fermions.  In this case, it is straightforward to construct the DW transfer matix in the $t$-direction following the approach of L\"uscher\cite{Luscher:1976ms}, and the analysis is similar to the construction of the DW transfer matrix in the $s$ direction\cite{Narayanan:1993sk,Furman:1994ky}. For simplicity, we will first consider the case of free fermions in one spatial dimension plus $s$ plus $t$ and then examine lattice calculations of correlation functions in three spatial dimensions plus  $s$ plus $t$ for both free fermions and fermions propagating in HYP smeared MILC configurations for 2+1 flavors of dynamical fermions.  We will show that free fermions and fermions in dynamical gluon configurations display the same oscillatory behavior, and implement a simple procedure for subtracting the oscillatory contributions from correlation functions.

\section{Domain wall transfer matrix in t}

We write the temporal gauge domain wall action in one spatial dimension in the following form:
\begin{eqnarray*}
%\begin{gathered}
D_{DW}(t,\vec{x},s;t',\vec{x'},s^\prime) &=& 
-\left( \frac{1+\gamma_t}{2}   \delta_{ t+1,t' } +
 \frac{1-\gamma_t}{2}   \delta_{ t-1,t' }   \right) \delta_{ \vec x,s; 
\vec x'  \!,s'  }
+\delta_{t,t'}
 D_{\{xs\}}(\vec{x},s;\vec x',s^\prime) , \\
D_{\{xs\}}(\vec{x},s;\vec x',s^\prime) &=& \left(\begin{array}{cc} B &	C \\
	-C^\dag & B
	\end{array}\right),\\
	\end{eqnarray*}
where
\begin{eqnarray*}
B_t(v,v^\prime)& = &(D+1-M)\delta_{v,v^\prime}
  - {1\over2}\sum_{\mu\in\{\vec{\hat x},\hat{s}\}}\left(
    U_{z,\mu}\delta_{v+\mu,v^\prime} + U_{z-\hat\mu,\mu}^\dag\delta_{v-\mu,v^\prime}
  \right),\\
C_t(v,v^\prime) &= &-{1\over2}\sum_{\mu\in\{\vec{\hat x},\hat{s}\}}\left(
    U_{z,\mu}\delta_{v+\mu,v^\prime} - U_{z-\hat\mu,\mu}^\dag\delta_{v-\mu,v^\prime}
  \right)\kappa_\mu,\\
U_{z,s,\mu} &=& 
\left\{ \begin{array}{cc}
  U_{z,\mu}, & \mu\in\{\hat{t},\vec{\hat{x}}\},\\
  -m, & \mu = \hat{s},\; s= 0 \,\,\,\mbox{or}\,\,\, L_s-1, \\
 1, & \mathrm{otherwise}.
	\end{array}\right\}  ,\\[.3cm]
v = (\vec{x},s), & & z= (t,\vec{x}), \quad \kappa = (1,-i), \quad D = 2. \\
%B &= &D+1 - M -\underbrace{\sum_{i=1}^{D-1}\cos k_i}_{\le D-1} 
%-\underbrace{\cos k_s}_{\le 1}
%\end{gathered}
\end{eqnarray*}
The action has the same form in three spatial dimensions, with $\kappa_\mu$ replaced by Pauli matrices $ \sigma_\mu$ for $\mu=\hat x, \hat y, \hat z$  and $1_{2\times2}$ for $\mu=\hat s$,  $B$  multiplied by $1_{2\times2}$, and D = 4.   In the free case, the eigenfunctions are plane waves and  $B$ has the value 
$$
B = D+1 - M -\underbrace{\sum_{i=1}^{D-1}\cos k_i}_{\le D-1} 
-\underbrace{\cos k_s}_{\le 1}
$$
Thus, if $M \le 1$, $B$ is nonnegative, whereas if  $M > 1$, there are values of $k$ such that $B$ is negative.
%%\begin{eqnarray*}
%%\begin{gathered}
%D_{DW}(t,\vec{x},s;t',\vec{x'},s^\prime) &=& 
%-\left( \frac{1+\gamma_t}{2}   \delta_{ t+1,t' } +
% \frac{1-\gamma_t}{2}   \delta_{ t-1,t' }   \right) \delta_{ \vec x,s; 
%\vec x'  \!,s'  }
%+\delta_{t,t'}
% D_{\{xs\}}(\vec{x},s;\vec x',s^\prime) , \\
%D_{\{xs\}}(\vec{x},s;\vec x',s^\prime) &=& \left(\begin{array}{cc} B &	C \\
%	-C^\dag & B
%	\end{array}\right),\\
%B_t(v,v^\prime)& = &(D+1-M)\delta_{v,v^\prime}
%  - {1\over2}\sum_{\mu\in\{\vec{\hat x},\hat{s}\}}\left(
%    U_{z,\mu}\delta_{v+\mu,v^\prime} + U_{z-\hat\mu,\mu}^\dag\delta_{v-\mu,v^\prime}
%  \right),\\
%C_t(v,v^\prime) &= &-{1\over2}\sum_{\mu\in\{\vec{\hat x},\hat{s}\}}\left(
%    U_{z,\mu}\delta_{v+\mu,v^\prime} - U_{z-\hat\mu,\mu}^\dag\delta_{v-\mu,v^\prime}
%  \right)\kappa_\mu,\\
%U_{z,s,\mu} &=& 
%\left\{ \begin{array}{cc}
%  U_{z,\mu}, & \mu\in\{\hat{t},\vec{\hat{x}}\},\\
%  -m, & \mu = \hat{s},\; s=L_s-1, \\
% 1, & \mathrm{otherwise}.
%	\end{array}\right\}  ,\\[.3cm]
%v = (\vec{x},s) & & z= (t,\vec{x}), \quad \kappa = (1,-i), \\
%B &= &D+1 - M -\underbrace{\sum_{i=1}^{D-1}\cos k_i}_{\le D-1} 
%-\underbrace{\cos k_s}_{\le 1}
%%\end{gathered}
%\end{eqnarray*}
%

A simple way to see the correspondence between the transfer matrix and the standard form of the domain wall action is to write the coherent  state path integral arising from a one-body  Hamiltonian and compare it with the domain wall path integral.  We use the following properties of fermion coherent states $| \xi \rangle$:
\begin{equation*}
a_\alpha |\xi\rangle = \xi_\alpha |\xi\rangle , \quad 1 = \int d\xi^*d\xi
 e^{-\xi_\alpha \xi_\alpha} |\xi\rangle \langle \xi | , \quad 
 \langle \xi |  e^{ a^\dag_\alpha H_{\alpha \beta} a_\beta }  |\xi\rangle
=e^{ \xi^*_\alpha \left[ e^H\right]_{\alpha \beta} \xi_\beta  }
\end{equation*}
Breaking evolution in Euclidean time into steps of size $\epsilon$ in the usual way and combining the terms indicated by the braces, we obtain
%
%\begin{equation*}
%\prod_t \int \! \!d\xi_t^*d\xi_t
%\end{equation*}
%
%CS path int 1'
%\begin{equation*}
%e^{-h}  =  \cdots   e^{-\epsilon a^\dag h a}   \hspace{2.3cm}  e^{-\epsilon a^\dag h a}     \hspace{4.5cm} e^{-\epsilon a^\dag h a}    \cdots
%\end{equation*}
%
%CS path int 1'
\begin{eqnarray*}
e^{-HT}  & = & \prod_t \int \! \!d\xi_t^*d\xi_t \cdots   e^{-\epsilon a^\dag H a} 
\underbrace{ e^{-\xi^*_{t \alpha} \xi_{t \alpha}}}
 |\xi_{(t)} \rangle
 \underbrace { \langle \xi _{(t)}| e^{-\epsilon a^\dag H a} }
    e^{-\xi^*_{(t-1) \alpha} \xi_{(t-1) \alpha}}
    \underbrace{ |\xi_{(t-1)} \rangle }
     \langle \xi _{(t-1)}| e^{-\epsilon a^\dag H a}    \cdots \\
 & = & \prod_t \int \! \!d\xi_t^*d\xi_t \cdots   e^{-\epsilon a^\dag H a}   |\xi_{(t)} \rangle
  \underbrace{ e^{-\xi^*_{t \alpha} \xi_{t \alpha}}e^{\xi^*_{t \alpha}  \left[ e^{-\epsilon H}
\right]_{\alpha \beta} \xi_{(t-1)\beta} } }
e^{-\xi^*_{(t-1) \alpha} \xi_{(t-1) \alpha}}
 \langle \xi _{(t-1)}| e^{-\epsilon a^\dag H a}    \cdots \\
\end{eqnarray*}
%
%CS complete 2'
%
%\begin{equation*}
% e^{-\xi_{(t-1) \alpha} \xi_{(t-1) \alpha}} |\xi_{(t-1)} \rangle \langle \xi _{(t-1)}|
%\end{equation*}
%
%CSpath int 2
%
%\begin{equation*}
%e^{-hT}  = \prod_t \int \! \!d\xi_t^*d\xi_t  \hspace{2cm} \cdots   e^{-\xi_{t \alpha} \xi_{t \alpha}}e^{\xi^*_{t \alpha}  \left[ e^{-\epsilon h}
%\right]_{\alpha \beta} \xi_{(t-1)\beta} }  
%\end{equation*}
%
so that the general term in the action connecting times $t$ and $t-1$ is
\begin{equation*}
S =  \cdots  -\xi^*_{t \alpha} \xi_{t \alpha}  + \xi^*_{t \alpha}  \left[ e^{-\epsilon H}
\right]_{\alpha \beta} \xi_{(t-1)\beta }.
\end{equation*}
Suppressing details for the lower components for brevity, for the upper components, 
the general term in the lattice action connecting times $t$ and $t-1$ can be brought into this form by writing
\begin{eqnarray*}
-S_F &= & \cdots -\psi^*_{t \alpha} B_{\alpha \beta}\psi_{t \beta}  + \psi^*_{t \alpha} M_{t \alpha (t-1)\beta} \psi_{(t-1)\beta} \\
&=&  \cdots  -\psi^*B^{1/2} B^{1/2} \psi + \psi^*  B^{1/2} B^{-1/2}M
 B^{-1/2} B^{1/2}\psi \\
& = & \cdots  -\underbrace{\psi^*B^{1/2}}_{\xi^*} \underbrace{B^{1/2} \psi }_\xi+ \underbrace{\psi^*  B^{1/2}}_{\xi^*}\underbrace{ B^{-1/2}M
 B^{-1/2}}_{e^{-\epsilon H} }\underbrace{B^{1/2}\psi}_{\xi} , \\
\end{eqnarray*}
%%\begin{eqnarray*}
%-S_F &= & \cdots -\psi_{t \alpha} B_{\alpha \beta}\psi_{t \beta}  + \psi^*_{t \alpha} M_{t \alpha (t-1)\beta} \psi_{(t-1)\beta} \\
%&=&  \cdots  -\psi^*B^{1/2} B^{1/2} \psi + \psi^*  B^{1/2} B^{-1/2}M
% B^{-1/2} B^{1/2}\psi
%\end{eqnarray*}
%
%\begin{equation*}
%-S_F =  \cdots -\psi_{t \alpha} B_{\alpha \beta}\psi_{t \beta}  + \psi^*_{t \alpha} M_{t \alpha (t-1)\beta} \psi_{(t-1)\beta}, 
%\end{equation*}
 transforming Grassman variables $\xi = B^{\frac{1}{2}} \psi$, and identifying $e^{-\epsilon H} =  B^{-\frac{1}{2}} M  B^{-\frac{1}{2}}$. 

Explicitly constructing the transfer matrix in this way (with appropriate treatment of antiparticle fields as well) for the domain wall action with one spatial dimension yields
\begin{eqnarray*}
%\begin{equation*}
%\begin{gathered}
e^{-H} &=&  \begin{bmatrix}
    1& \\
    & { B_1^{-1}}
  \end{bmatrix}
\cdot
  \begin{bmatrix}
    1 & C_0 \\
    C_1^\dag & 1+C_1^\dag C_0 
  \end{bmatrix}
\cdot\begin{bmatrix}
    B_0 & \\
    & { 1}
  \end{bmatrix}  \\ [.3cm]
Z &=& \mathrm{Tr}\left[e^{-\hat a^\dag H \hat a}\right]^N.
 %\end{gathered}
%\end{equation*}
\end{eqnarray*}
%
%Hence, to isolate the oscillatory terms in the transfer matrix for a free particle in three dimensions $( x, t, s)$ we calculate the eigenmodes 
Hence, to isolate the oscillatory terms in the transfer matrix, we calculate the eigenmodes 
\begin{equation}
e^{-H} \psi(x,s) = \eta \psi(x,s). \label{eigen}
\end{equation}
The eigenfunctions in Eq.~\ref{eigen} are orthogonal with the weight matrix
%$\begin{pmatrix}
%B & 0 \\ 
%0 & B
%\end{pmatrix}$
%and
$\left(\begin{smallmatrix}
B & 0 \\ 
0 & B
\end{smallmatrix}\right)$
 and all the eigenvalues are necessarily real. 
Negative eigenvalues $\eta$ correspond to terms that change sign under evolution by one time step and thus produce oscillatory behavior in temporal correlation functions.

\section{Negative eigenmodes for free fermions in 3 dimensions}

\begin{figure}[t]
\begin{center}
\hbox{ \hspace{-.8cm}\raisebox{0cm}{\includegraphics[width=17pc,angle=0,scale=1.2]{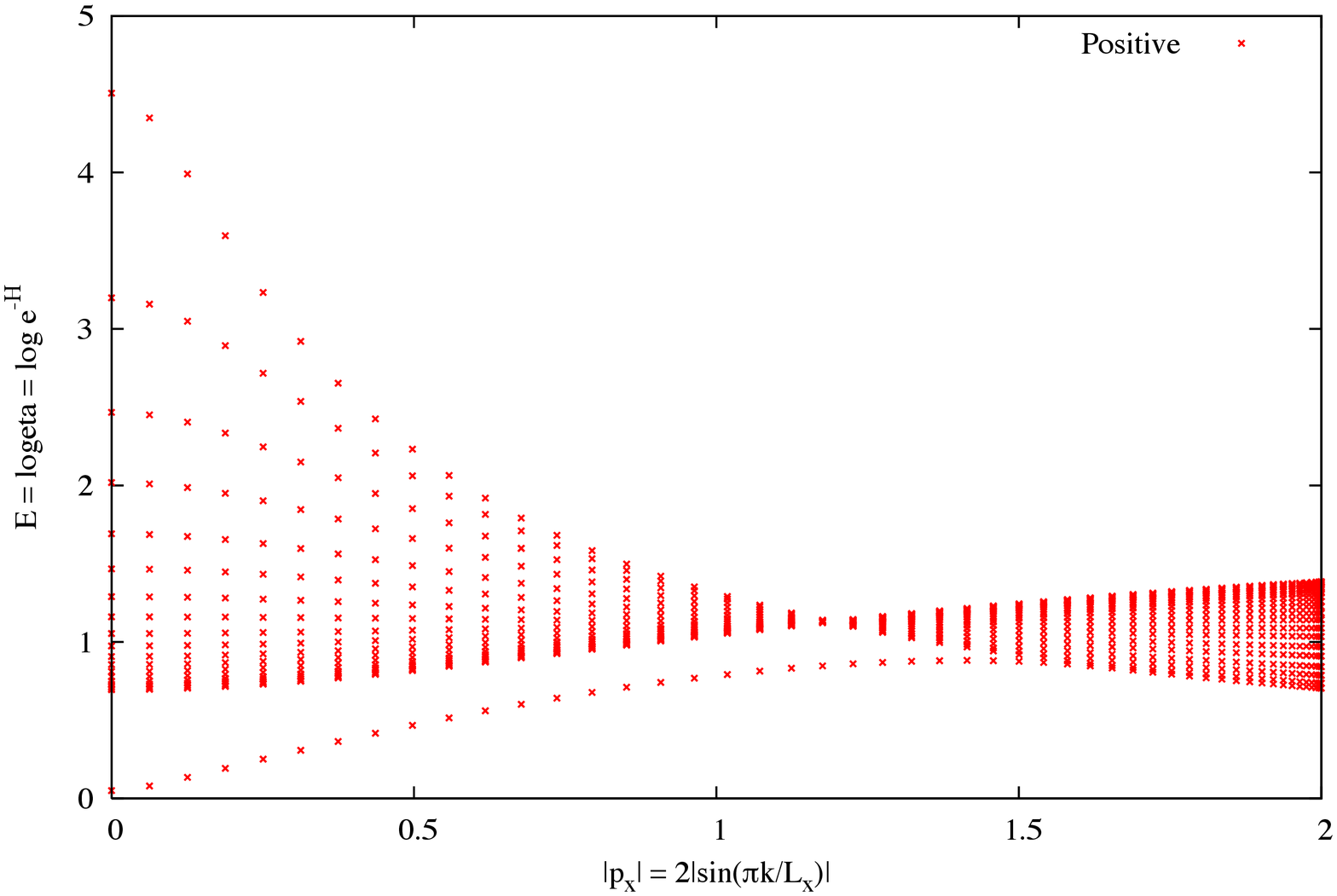}}
\hspace{-1cm}\includegraphics[width=17pc,angle=0,scale=1.2]{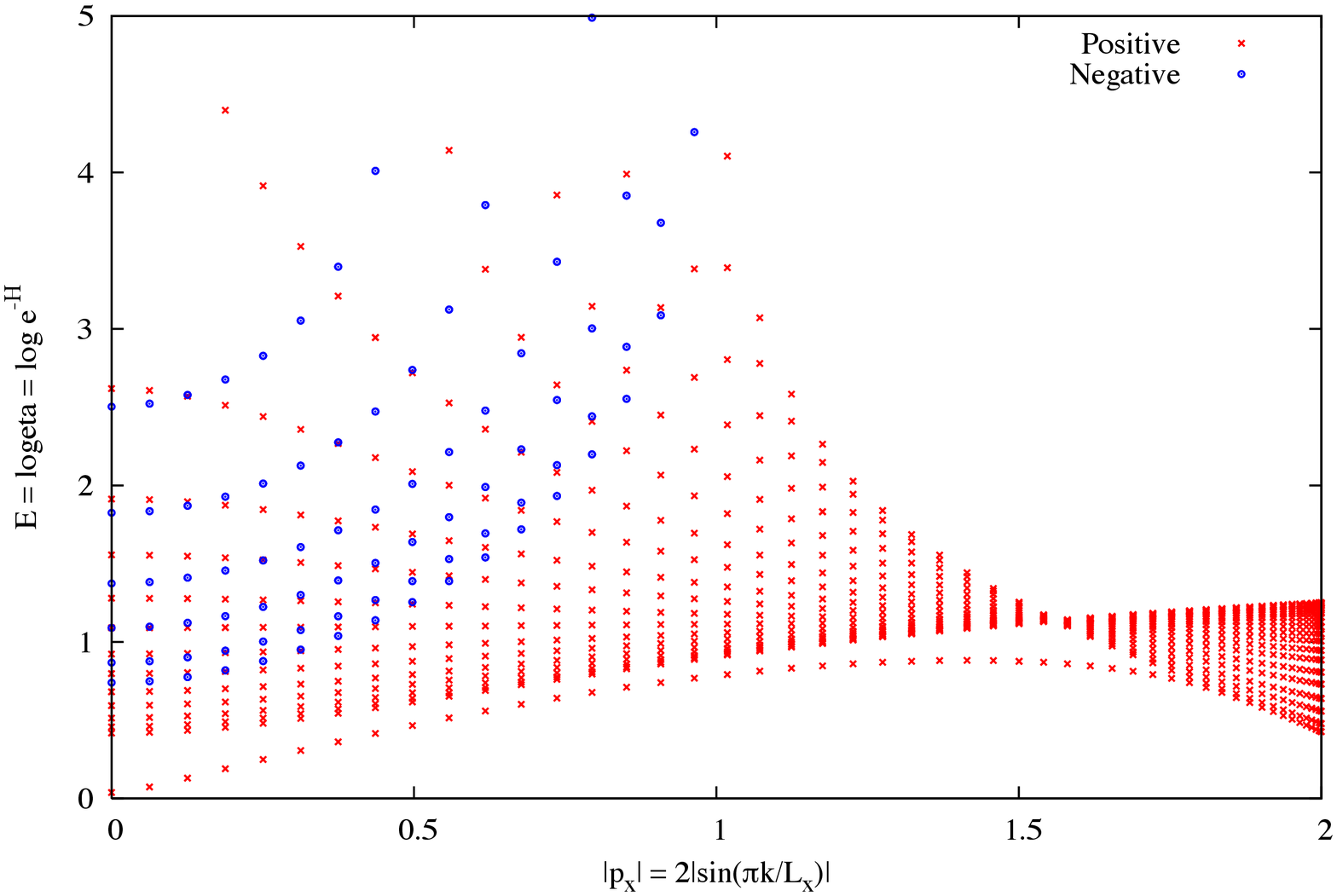}}
\vspace{-.7cm}
\hbox{ \hspace{-.8cm}\raisebox{0cm}{\includegraphics[width=17pc,angle=0,scale=1.2]{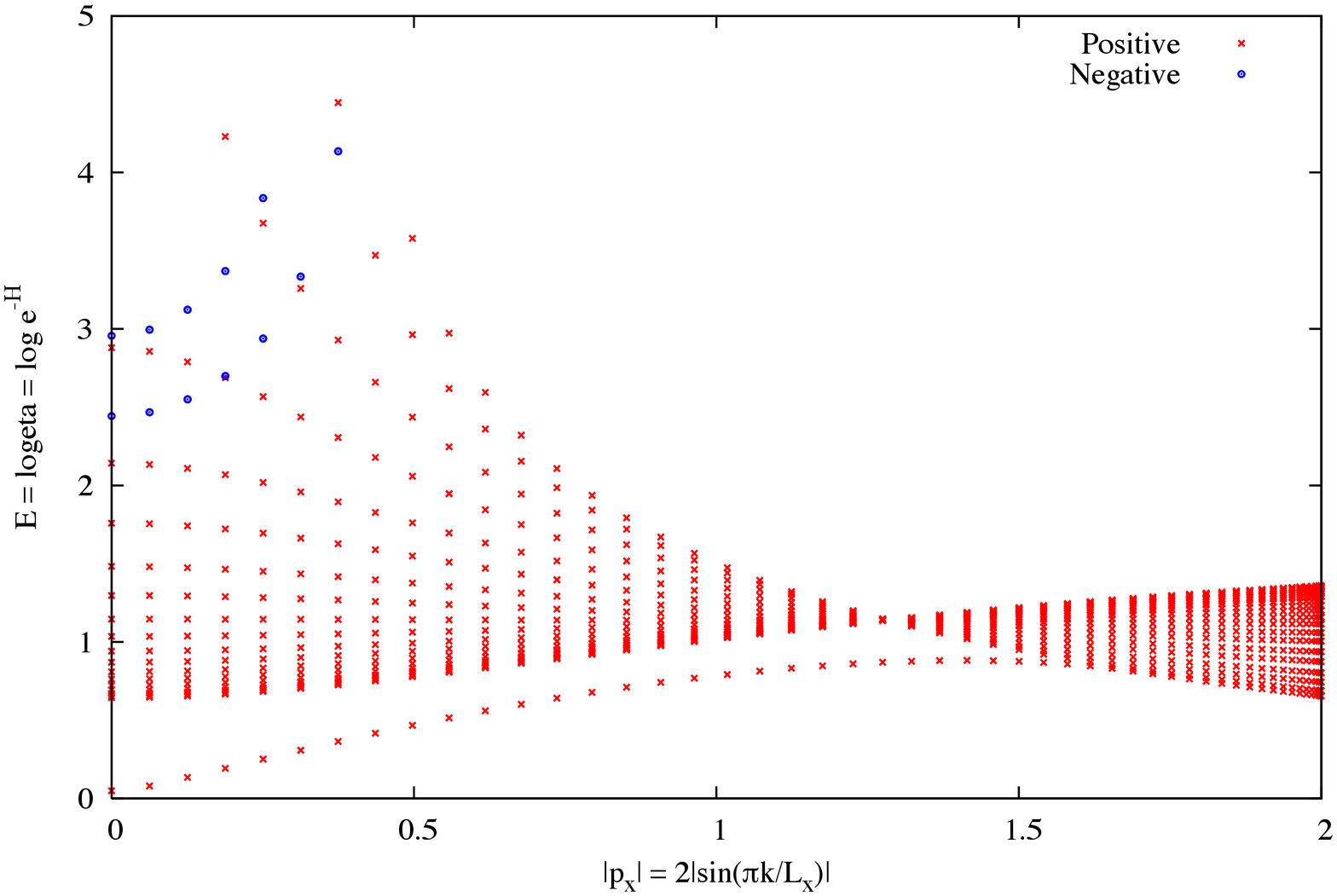}}
\hspace{-1cm}\includegraphics[width=17pc,angle=0,scale=1.2]{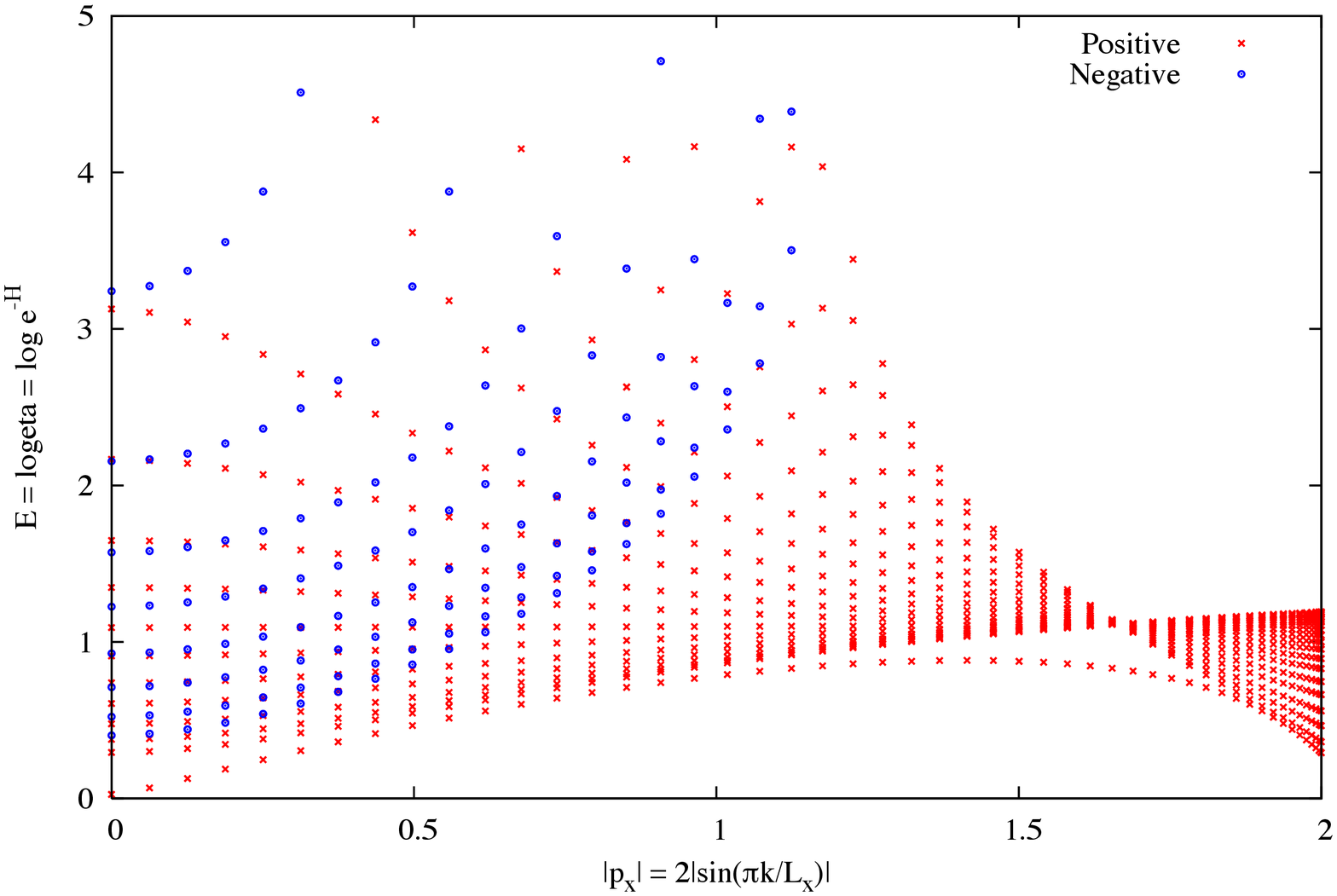}}
\vspace{-1.8cm}
\hbox{ \hspace{-.8cm}\raisebox{.4cm}{\includegraphics[width=17pc,angle=0,scale=1.2]{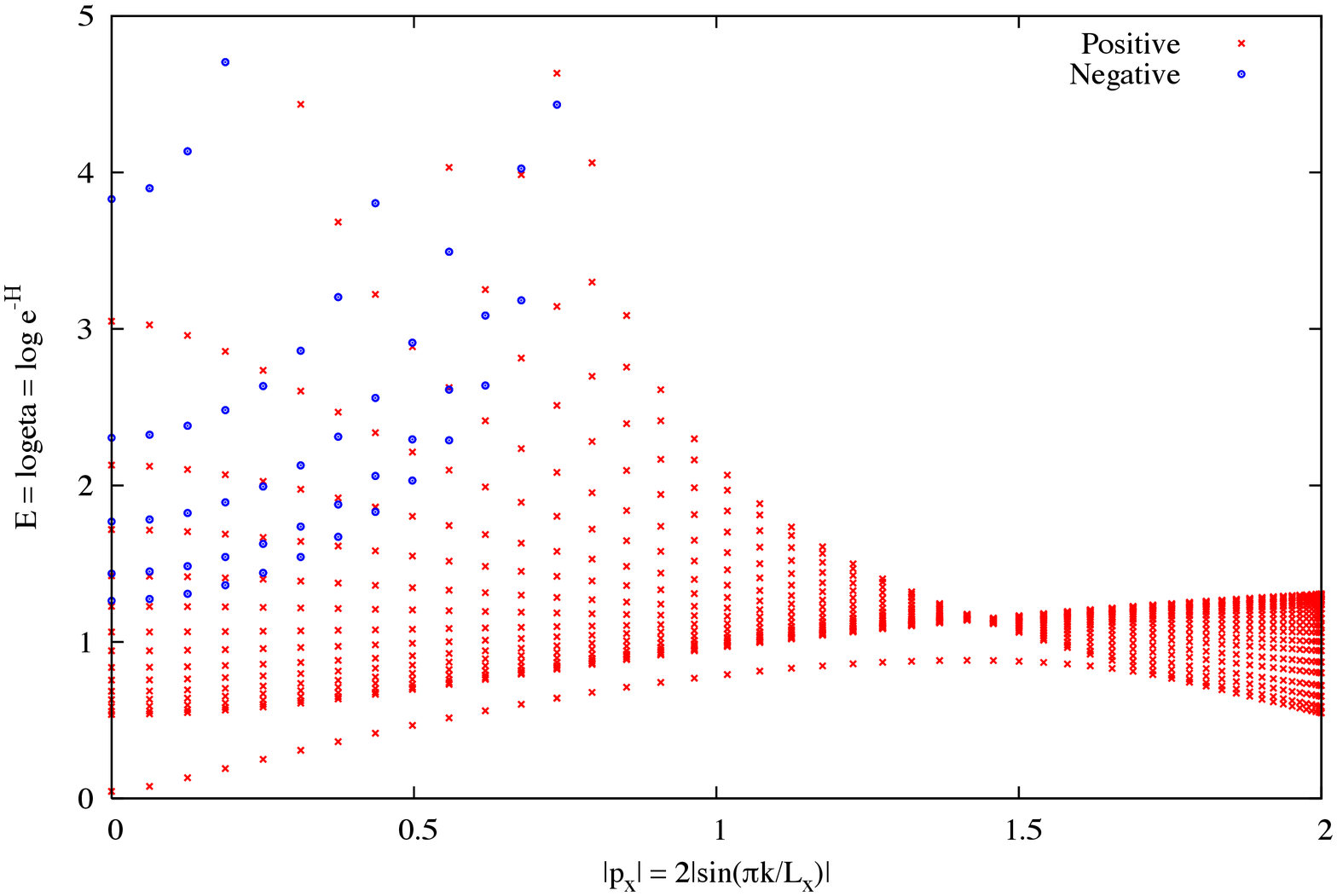}}
\hspace{-1.1cm}\includegraphics[width=17pc,angle=0,scale=1.3]{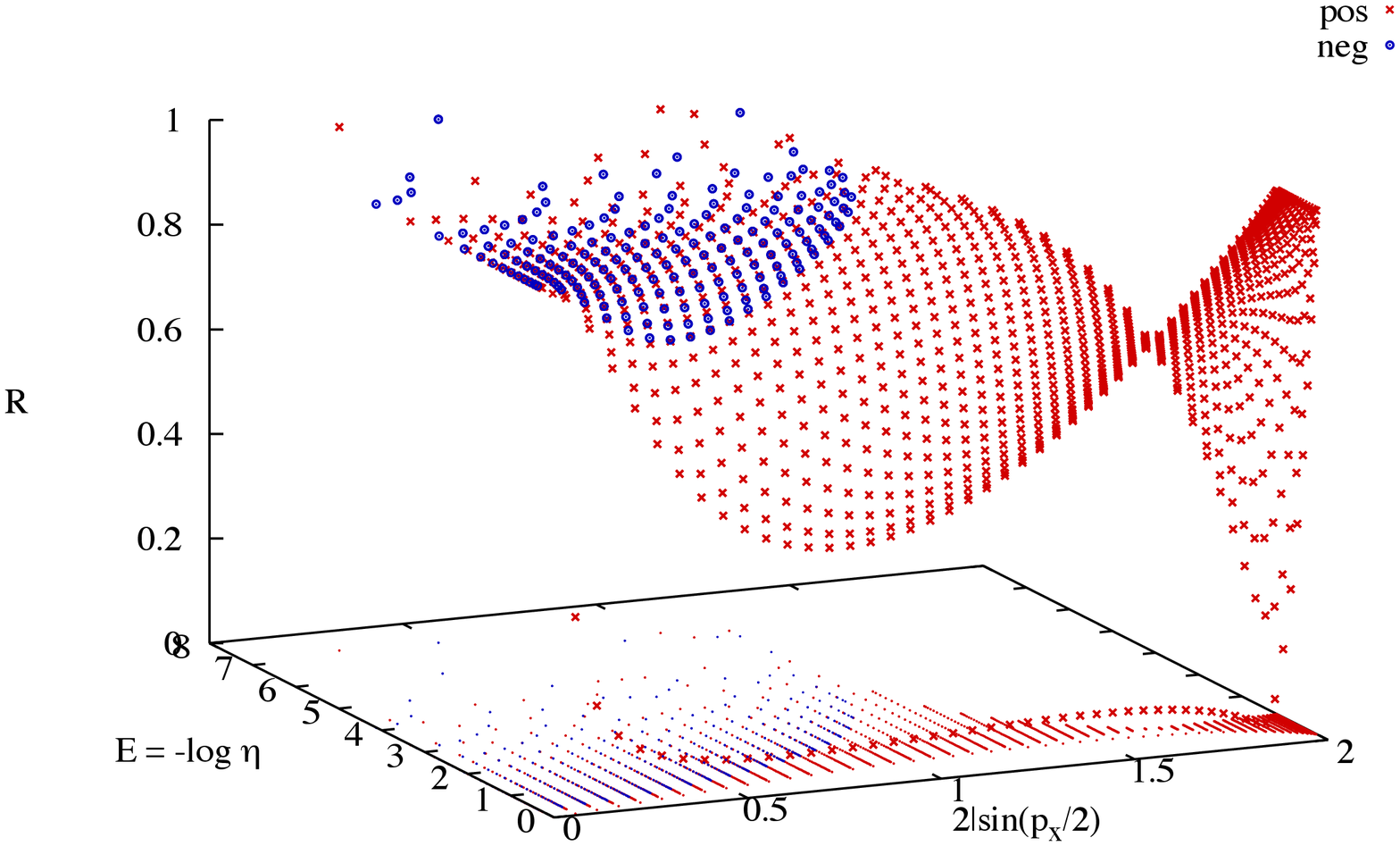}}
\vspace{-.9cm}
\caption{\label{oned}  $E= log|\eta|$ as a function of  $p$ for positive and negative eigenvalues $\eta$ for free fermions in one spatial dimension. The left plots show $M$ = 1, 1.1, and 1.3 and the top right plots show M =  1.5 and 1.7. The lower right plot shows the chirality $R$ for all the modes at $M$ = 1.7. }
\end{center}
\vspace{-1.2cm}
\end{figure}

The eigenvalues and eigenfunctions in Eq.~\ref{eigen} were computed for free fermions in one space plus time and flavor dimensions on a mesh with $N_x$ = 100,  $N_s$ = 20 and $ma=0.05$,  and typical results are shown in Fig.~\ref{oned}.   The eigenfunctions are momentum eigenstates and the quantity  $E= log|\eta|$  is plotted   as a function of  $p$ for both positive  eigenvalues $\eta$ (denoted by red crosses), where it corresponds to the energy of the mode, and negative eigenvalues (denoted by blue circles).  At $M$ = 1, there are only positive modes, as expected, and there is a single physical mode corresponding to the free quark dispersion relation.  As M increases  through 1.1, 1.3, 1.5, and 1.7, the blue points show the increasing presence of negative eigenmodes.  However, all these modes have energies comparable to the positive modes at or above the cutoff scale, and are thus expected to simply be additional cutoff effects.  The lower right plot shows the chirality  $R = \frac{\psi^\dag_R \psi_L +    \psi^\dag_L \psi_R  }
{\psi^\dag_R \psi_R +    \psi^\dag_L \psi_L  }$,  which should be small for chiral modes, for each of the points shown in the middle right plot  for $M=1.7$.  As expected, only the free mode is chiral, and the negative modes at the cutoff scale behave the same as the positive modes at that scale. Thus all the evidence suggests that these modes are innocuous and can be treated like other cutoff effects.

\section{Oscillations in correlation functions in 5 dimensions}

\begin{figure}[tb]
\begin{center}
\vspace{-.5cm}
\hbox{ \hspace{-.8cm}\raisebox{0cm}{\includegraphics[width=17pc,angle=0,scale=1.2]{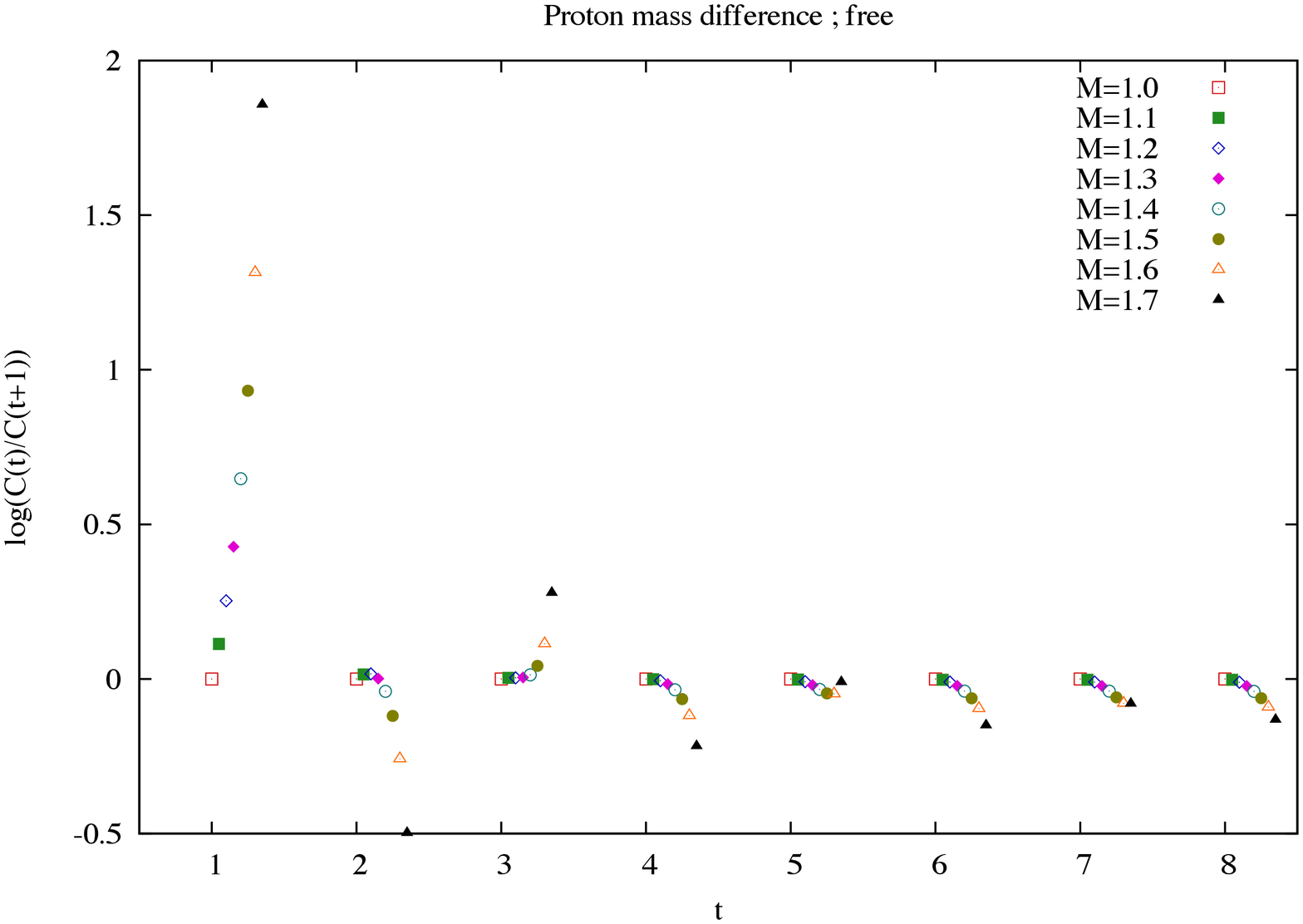}}
\hspace{-1cm}\includegraphics[width=17pc,angle=0,scale=1.2]{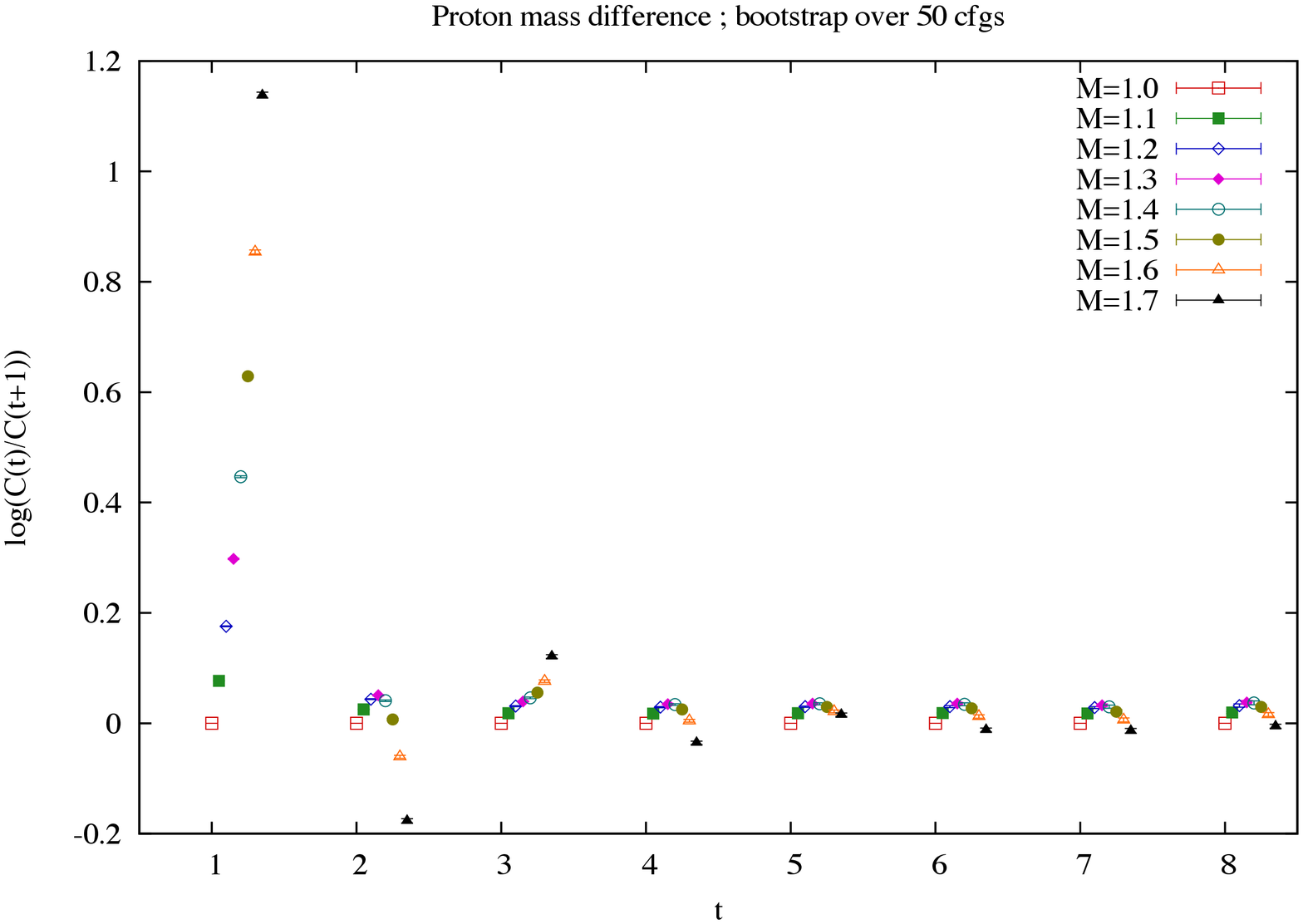}}
\vspace{-1cm}
\caption{\label{effm}  Proton effective mass $E_{eff} = log \,C(t)/C(t+1)$ in three space dimensions for free fermions (left) and HYP smeared MILC lattices (right)  for domain wall masses $1 \le M \le 1.7$. To display the oscillations, the plot shows  $E_{eff} (M) - E_{eff} (1)$.}
\end{center}
\end{figure}

Analogous negative modes arise for free fermions in three space plus time and flavor dimensions, and their effect on the proton two-point correlation function $C_2(t)$   is shown in the left panel of  Fig.~\ref{effm} for a  $20^3 \times 32 $ lattice with $m a= 0.0810$ .   To display the oscillations, we calculate the effective mass   $E_{eff} = log C(t)/C(t+1)$ and plot the difference between the effective masses at  $M$ and $ M = 1$,  $E_{eff} (M) - E_{eff} (1)$. For clarity, the results at each $t$ are displaced slightly to the right with increasing $M$ to avoid overlapping symbols and to indicate the behavior with increasing $M$.  Note that oscillations first become visible at $M = 1.4$ and increase strongly as $M$ is  increased to 1.7, rising on odd time slices and falling on  even time slices. 

The same qualitative behavior is observed in the right panel of  Fig.~\ref{effm} for domain wall fermions on HYP-smeared MILC lattices, for which the chopped lattice size is $20^3 \times 32$, $a$ = 0.125 fm, $m_\pi$ = 759 MeV and 50 configurations were used.  For the MILC lattices, the oscillations become visible at  $M = 1.5$ and grow much larger by $M = 1.7$.

\begin{figure}[tb]
\vspace{-.5cm}
\begin{center}
\hbox{ \hspace{-.15cm}\raisebox{0cm}{\includegraphics[width=17pc,angle=-90,scale=0.74]{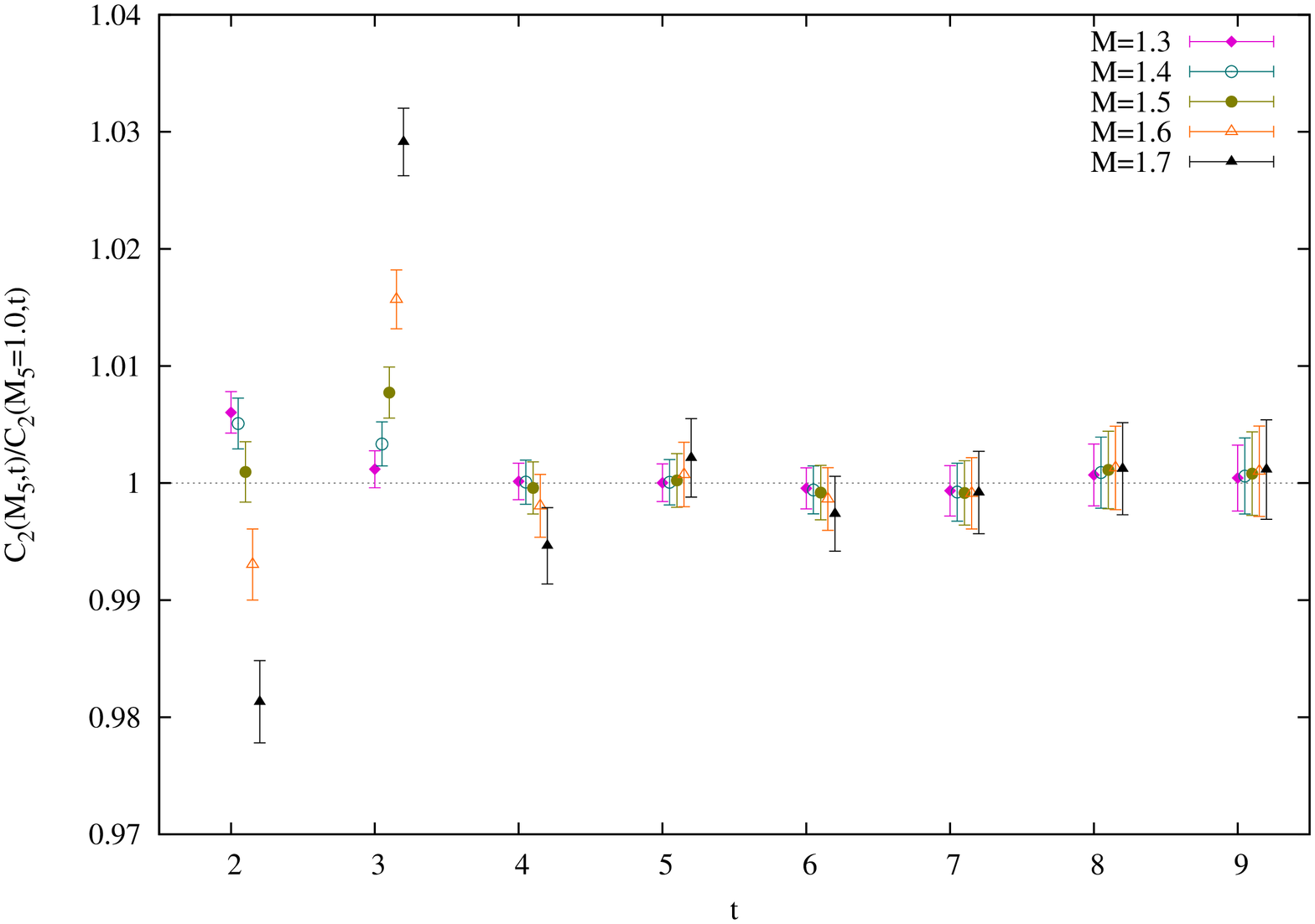}}
\hspace{0cm}\includegraphics[width=17pc,angle=-90,scale=0.74]{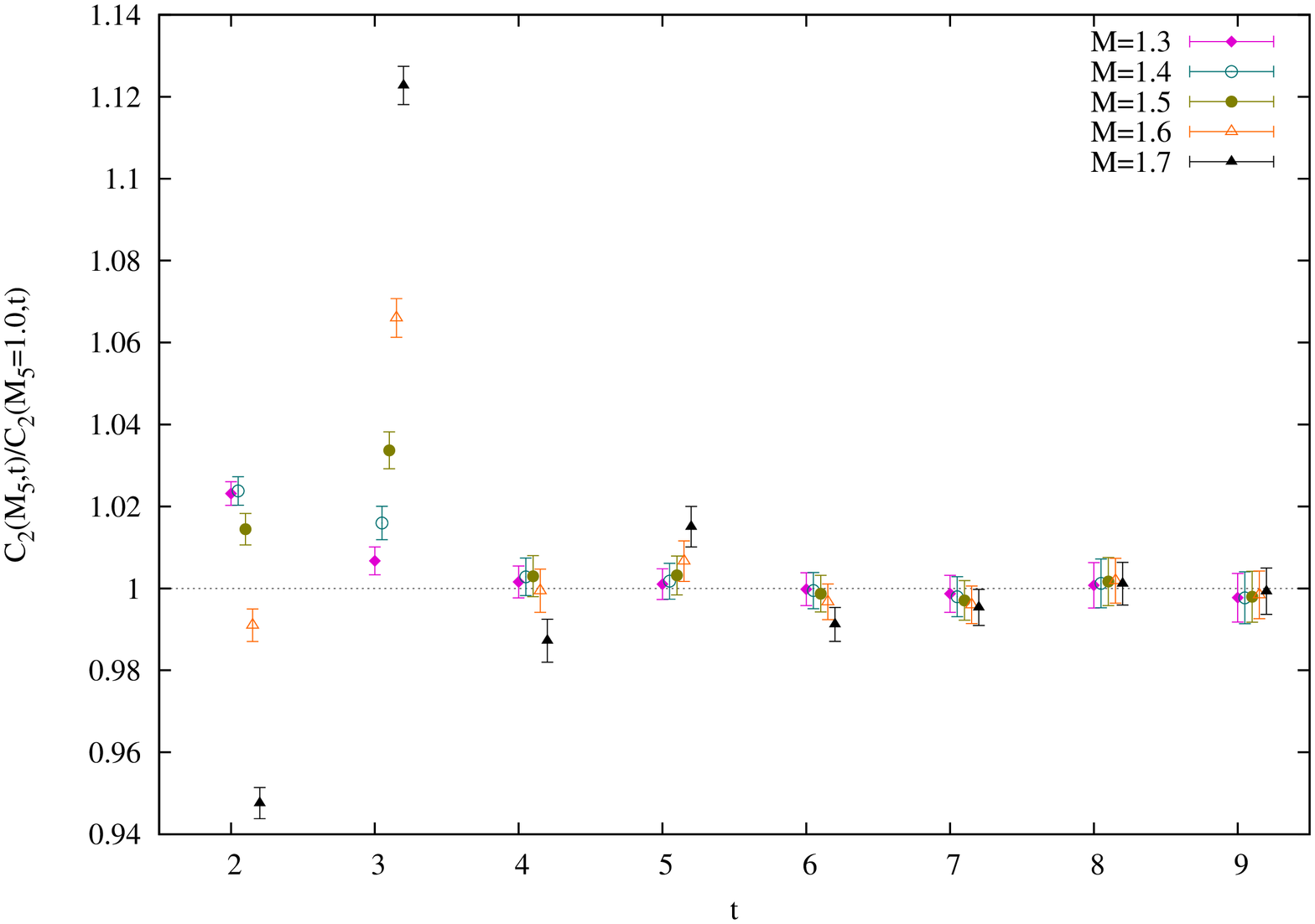}}
\vspace{-.3cm}
\caption{\label{ratio}  Correlation function ratio $R = C_2(M, t) /C_2(1,t) e^{\Delta m t }$ for the pion (left) and nucleon (right) for HYP smeared MILC lattices  }
\vspace{-.5cm}
\end{center}
\end{figure}

An alternative way of detecting oscillations, which is  convenient computationally, is to calculate the ratio  of the correlation function at  $M$ to that at $M = 1$, $R = C_2(M, t) /C_2(1,t) $.  This ratio is shown in Fig.~\ref{ratio} for the pion and the proton, where for aesthetic purposes, we have also multiplied by $e^{\Delta m t }$ to make the curve flat.  Note that oscillations  become visible in both the pion and proton at $M= 1.5$, and become large by $M= 1.7$.   Also note that the effect is much larger for the proton, where  for $M= 1.7$ it is 12\% whereas for the pion it is 3\%. 

\section{Subtraction of oscillating terms}

\begin{figure}[tb]
\begin{center}
\vspace{-.4cm}
\hbox{ \hspace{-1.7cm}\raisebox{.24cm}{\includegraphics[width=17pc,angle=0,scale=1.4]{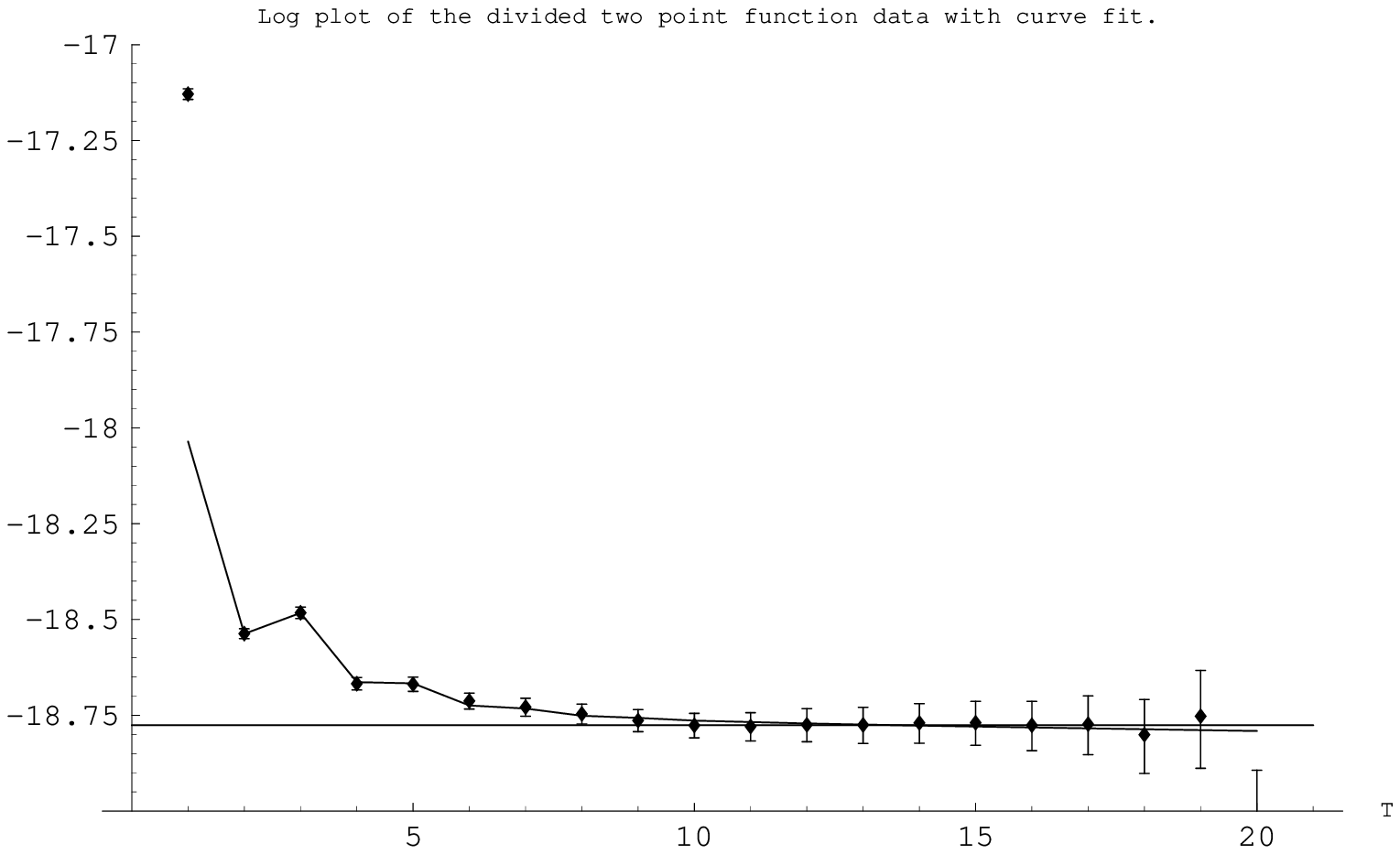}}
\hspace{-1cm}\includegraphics[width=17pc,angle=0,scale=1.05]{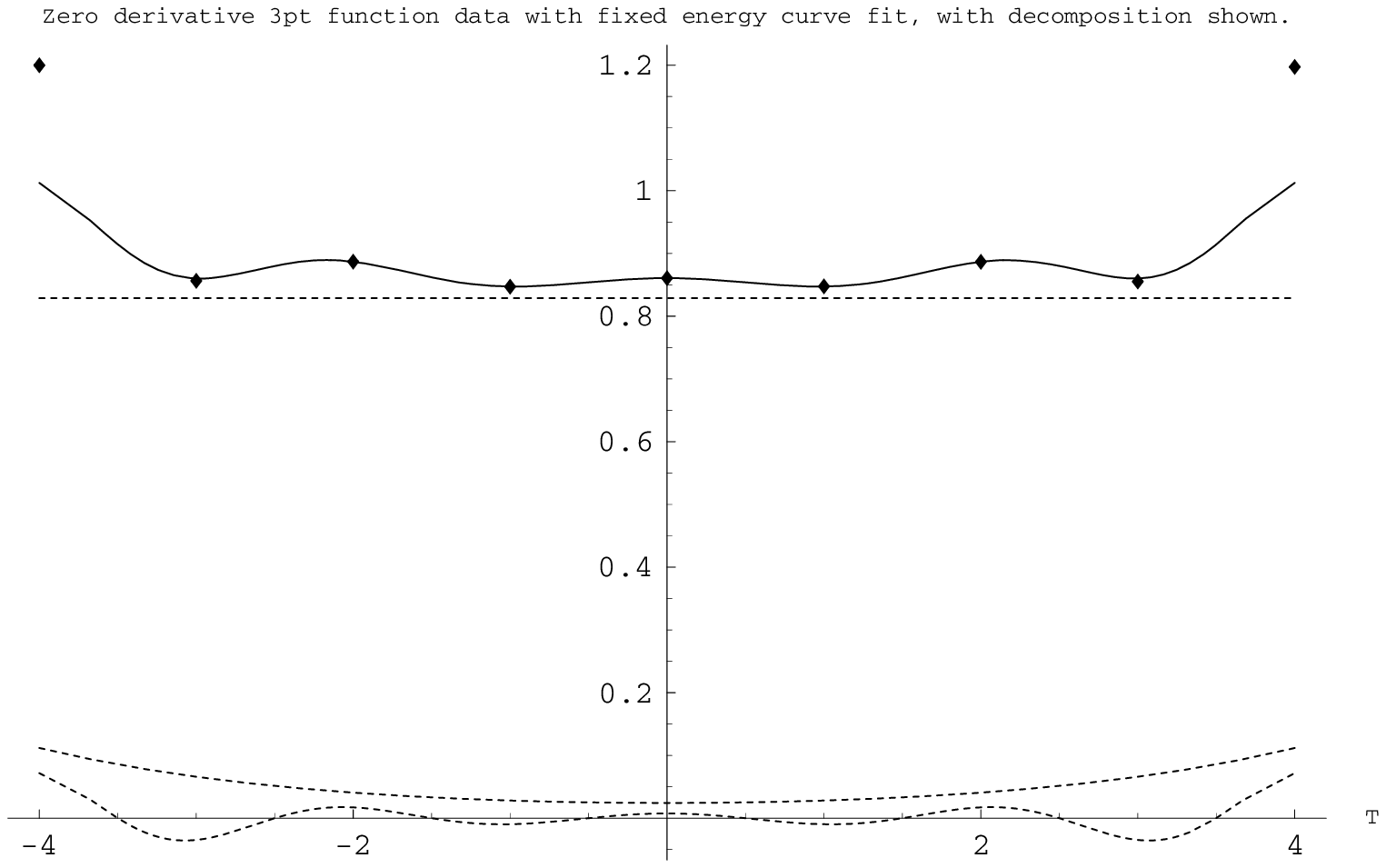}}
\vspace{-.5cm}
\caption{\label{fit}  Six-parameter fit to the nucleon two-point  function (left) and three-parameter fit to the normalized  vector current three-point function (right) for DW fermions on HYP-smeared MILC lattices at $M$ = 1.7. }
\vspace{- 1cm}
\end{center}
\end{figure}

Since all the evidence suggests that the negative eigenmodes are innocuous excitations at the cutoff scale for both free quarks and QCD,  a straightforward way to remove their effect from correlation functions is to fit the correlation function with the sum of an exponential for the physical ground state, 
exponentials for physical excited states, and exponentials multiplied by $(-1)^t$ for the negative eigenmodes of the transfer matrix.  Thus, for the  2-point function, including the leading corrections yields
\begin{equation*}
f(t) = A_0 e^{-m_0 t} + A_1 e^{- m_1 t} + \tilde A_1 e^{- \tilde m_1 t}(-1)^t ,
\end{equation*}
and for the normalized 
3-point function for a matrix element we obtain
\begin{equation*}
\frac{\langle J(\frac{T}{2} ) \, {\cal O}(t) \, J(  - \frac{T}{2}  )  \rangle   }{\langle J(\frac{T}{2} ) \,J(-\frac{T}{2} ) \rangle} =
B_0 + B_1 \cosh[(m_1-m_0) t] + \tilde B_1 \cosh[(\tilde m_1-m_0) t] (-1)^t .
\end{equation*}
Figure~\ref{fit}  shows the application of this analysis\cite{bratt}
for MILC lattices with $m_\pi$ = 759 MeV.  The 2-point function determines all three masses as well as three amplitudes, and using these masses, the normalized 3-point function determines the three amlitudes $B_0, B_1$, and $\tilde B_1$ and thereby enables accurate extraction of the physical matrix element, $B_0$,  for the vector current operator.

%
%two point function
%6 params   MILC  divided by $e^{-0.970t}$ 

%vector current matrix element  ratio of 3-pt fn to 2-point fn  3 paramters

%%
%\begin{figure}[tb]
%\begin{minipage}{17.5pc}
%\includegraphics[width=17pc,angle=0,scale=1]{Ratio_F2_F1new}
%\caption{\label{F2_F1}Isovector form factor ratio $F_2/F_1$ at three masses compared with experiment~\cite{Kelly:2004hm}.} % F2/F1
%\end{minipage}
%\hspace{0.5pc}
%\begin{minipage}{17.5pc}
%\raisebox{0cm}{\includegraphics[width=17pc,angle=0,scale=1.0]{ratio_GA_F1_u-d_vs_expt}}
%\caption{\label{GA_F1}Isovector form factor ratio $G_A/F_1$ at four masses compared with experiment.}%GA/F1
%\end{minipage}
%\end{figure}

%

%\begin{figure}[tb]
%\begin{center}
%\includegraphics[width=17pc,angle=0,scale=1.8]{A30_A10}\vspace{2pc}
%\caption{\label{A30_A10} Comparison of the ratio $ A_{30} / A_{10}$ for $u-d$ and $u+d$ at four masses with a phenomenological fit to generalized parton distributions. }% A30/A10
%\end{center}
%\end{figure}

\section{Conclusions}
In summary, we have calculated the transfer matrix in the Euclidean time direction for free domain wall fermions and  shown that negative eigenmodes arise for $M > 1$ and produce oscillatory behavior in lattice correlation functions.  We argue that these are innocuous lattice effects at the cutoff scale and show how their effects can be removed from QCD correlation functions.  In the future, however, since the additive mass shift depends on the gauge action and lattice spacing, we believe care should be taken in choosing the domain wall mass $M$  for each action and lattice spacing to eliminate the appearance of these modes in lattice calculations.

This work was supported in part by funds provided by the U.S. Department of Energy under grant DE-FG02-94ER40818 and computer resources were provided on the MIT Blue Gene computer.

\end{document}